\documentclass[twocolumn,english,aps,prl,showpacs]{revtex4}
\usepackage[T1]{fontenc}
\usepackage{amsmath,graphicx,amssymb,epsfig,babel,dsfont,color}

\begin{document}

\title{Fundamental Limits for Light Absorption and Scattering Induced by Cooperative Electromagnetic Coupling}

\author{Jean-Paul Hugonin$^{1}$, Mondher Besbes$^{1}$ and Philippe Ben-Abdallah$^{1,*}$}

\affiliation{$^1$Laboratoire Charles Fabry,UMR 8501, Institut d'Optique, CNRS, Universit\'{e} Paris-Sud 11,
2, Avenue Augustin Fresnel, 91127 Palaiseau Cedex, France.}
\email{pba@institutoptique.fr}

\date{\today}


\begin{abstract}
Absorption and scattering of electromagnetic waves by dielectric media are of fundamental importance in many branches of physics. Here we analytically derived the ultimate upper limits for the absorbed and scattered powers by any system of coupled particles and give sufficient conditions to reach these limits paving so a way for a rational design of optimal metamaterials.

\end{abstract}

\maketitle

Understanding light absorption and scattering mechanisms  is of prime importance in optical physics, photonics and plasmonics. Nowadays, the nanofabrication technologies provide tools to design artifical materials for engineering light-matter interactions. Plasmonic nanoparticles agregates are, for instance, artificial structures which have been subject of intense researchs \cite{Nordlander1,Nordlander2}. The collective interactions of localized plasmon modes can give rise to pronounced Fano resonances \cite{Fano1,Fano2} or lead to electromagnetic induced transparency \cite{EIT1} or induced absorption \cite{EIT2} phenomena. 

In this Rapid Communication, we analyze how strong light absorption and scattering can be in arbitrary sets of resonant particles when they are cooperatively coupled. This problem has been partly considered  in the specific case of  single objects \cite{Bohren, Fan1, Fan2, Fan3,Fleury,Alu,Grigoriev}, objects with simple symmetry shapes \cite{Khokhlov,Soljacic}, periodic arrays of dipoles \cite{Tretyakov} or dilute distibution of scatterers \cite{Joannopoulos}. Here, we provide  a general answer  to this problem by deriving, in the framework of the linear classical electrodynamics, the fundamental upper bounds for light absorption and scattering for arbitary distributions of resonant particles when potentially strong cooperative coupling mechanisms take place and when the field radiated by the multipolar orders of particles participate to the coupling.  We also propose a general inverse strategy to reach these limits.

Let us start, for pedagogical reasons, by considering  dipolar systems made with  N dipoles of  dipolar moment $\mathbf p_{i}$ spatially distributed at the position  $\mathbf{r}_i$ in a transparent medium of permittivity $\epsilon_h$ and higlighted by an incident field $\mathbf E_{inc}$. The local electric field $\mathbf{E}_{loc}$ measured at any point results from the superposition of external incident  and scattered fields. Therefore it takes the self-consistent form \cite{Purcell,Draine}
\begin {equation}
 \mathbf E_{loc}(\mathbf{r})=\mathbf E_{inc}(\mathbf{r})+\omega^2 \mu_0\underset{j}{\sum}\mathds{G}_0(\mathbf{r},\mathbf{r}_j)\mathbf p_{j}
,\label{Eq:external_field}
\end{equation}
where  $\mathds{G}_{\mathrm{0}}(\mathbf{r},\mathbf{r}')=\frac{\exp({\rm i}k\rho)}{4\pi \rho}\times
\left[\left(1+\frac{{\rm i}k\rho-1}{k^{2}\rho^{2}}\right)\mathds{1}+\frac{3-3ik\rho-k^{2}\rho^{2}}{k^{2}\rho^{2}}\widehat{\boldsymbol{\rho}}\otimes\widehat{\boldsymbol{\rho}}\right]$
is the free space Green tensor in the host material defined with the unit vector $\widehat{\boldsymbol{\rho}}\equiv\boldsymbol{\rho}/\rho$,
$\boldsymbol{\rho}=\mathbf{r}-\mathbf{r}'$, $k$ is the wavector in the host material while $\mathds{1}$
denotes the unit dyadic tensor. By introducing the vectorial fields $ \mathbf p=^t(\mathbf p(\mathbf{r_1}),...,\mathbf p(\mathbf{r_N}))$, $ \mathbf  {E}_{loc}=^t(\mathbf  {E}_{loc}(\mathbf{r_1}),...,\mathbf {E}_{loc}(\mathbf{r_N}))$, $ \mathbf  {E}_{inc}=^t(\mathbf  {E}_{inc}(\mathbf{r_1}),...,\mathbf  {E}_{inc}(\mathbf{r_N}))$ and using the Poynting theorem \cite{Jackson}, the total power dissipated in this system highligted by a monochromatic incident field writes \cite{Jackson, Ben-Abdallah1,Ben-Abdallah2} 
\begin {equation}
 \begin{split}\mathcal{P}_{abs}(\omega)=-\frac{\omega }{2} Im[\mathbf p^\dagger(\omega).\mathbf E_{loc}(\omega)].
\end{split} \label{Eq:power1}
\end{equation}
Using relation (\ref{Eq:external_field}) and the reciprocity principle \cite{Landau}, this expression  can be recasted in term of the incident field 
\begin {equation}
 \begin{split}\mathcal{P}_{abs}(\omega)=-\mathbf p^\dagger.(\frac{\omega^3\mu_0}{ 2} [Im\mathcal{G}]).\mathbf p
+Re[\mathbf p^\dagger.i\frac{\omega }{2}\mathbf E_{inc}].
\end{split} \label{Eq:power2}
\end{equation}
where we have set the $N\times N$ block matrix of component
$\mathcal{G}_{ij} =\mathds{G}_{0}(\mathbf{r_i},\mathbf{r_j})$. 
Hence, the maximal power than a set of dipoles can dissipate under an external highligting reads (see Supplemental Material \cite{SupplMat})
\begin {equation}
\mathcal{P}^{max}_{abs}(\omega)=\frac{1}{8\omega\mu_0} \mathbf E_{inc}^\dagger.[Im\mathcal{G}]^{-1}.\mathbf E_{inc}.
\label{Eq:absorb_max}
\end{equation}
This expression is the first result of the Letter. It represents the fundamental limit for light absorption by any system of dipoles which collectively interact.This expression depends only on the geometric configuration and on the characteristics of incident field. It immediately follows from it that the maximal absorption cross-section for a set of dipoles reads
\begin {equation}
\sigma_{abs}^{max}=\frac{c}{4\omega}\frac{\mathbf E_{inc}^\dagger.[Im\mathcal{G}]^{-1}.\mathbf E_{inc}}{\sqrt\epsilon_h|\mathbf E_{inc}|^2}.
\label{Eq:absorb_section}
\end{equation}
This result could seem at first view counter intuitive. Indeed, with resonant particles we could expect a blowup mechanism for light absorption  allowing to go beyond this limit. However, this intuition is wrong. To be convinced of this let us simply examine the case of an isolated particle of polarizability $\alpha=\alpha^{'}+i\alpha^{"}$. In this simple case, the dissipated power writes \cite{Tretyakov}
\begin {equation}
\mathcal{P}_{abs}(\alpha^{'},\alpha^{"})=\frac{\omega}{2\mu_0 c^2} \mid\mathbf E_{inc}\mid^2\{\alpha^{"}-\frac{\omega^3}{6\pi c^3}\mid\alpha\mid^2\}.
\label{Eq:particle}
\end{equation} 
Then, a straighforward calculation shows that the optimal polarizability is  $\alpha_{opt}=3i\pi\frac{c^3}{\omega^3}$ and it leads to the maximal dissipated power $\mathcal{P}^{max,sing}_{abs}(\omega)=\frac{3}{4} \frac{\pi c}{\omega^2\mu_0 n_h}\mid\mathbf E_{inc}\mid^2$  which exactly corresponds (using  $Im\mathcal{G} =\frac{k}{6\pi}\mathds{1}$) to the value predicted by relation (\ref{Eq:absorb_max}).

Similarly, we can maximize the power 
\begin{equation}
\mathcal{P}_{sct}=\mathbf p^\dagger. (\frac{\omega^3\mu_0}{ 2}[Im\mathcal{G}]).\mathbf p
\end{equation}
 scattered by a set of dipoles under the constraint $\mathcal{P}_{abs}\geq 0$ to find (see Supplemental Material \cite{SupplMat})
\begin {equation}
\mathcal{P}^{max}_{sct}=\frac{1}{2\omega\mu_0} \mathbf E_{inc}^\dagger.[Im\mathcal{G}]^{-1}.\mathbf E_{inc}.
\label{Eq:scatter_max}
\end{equation}
Comparing this expression to relation (\ref{Eq:absorb_max}) we see that the maximal power than a set of dipoles can scatter is four times equal to the maximal power it can absorb.

\begin{figure}[Hhbt]
\includegraphics[scale=0.34]{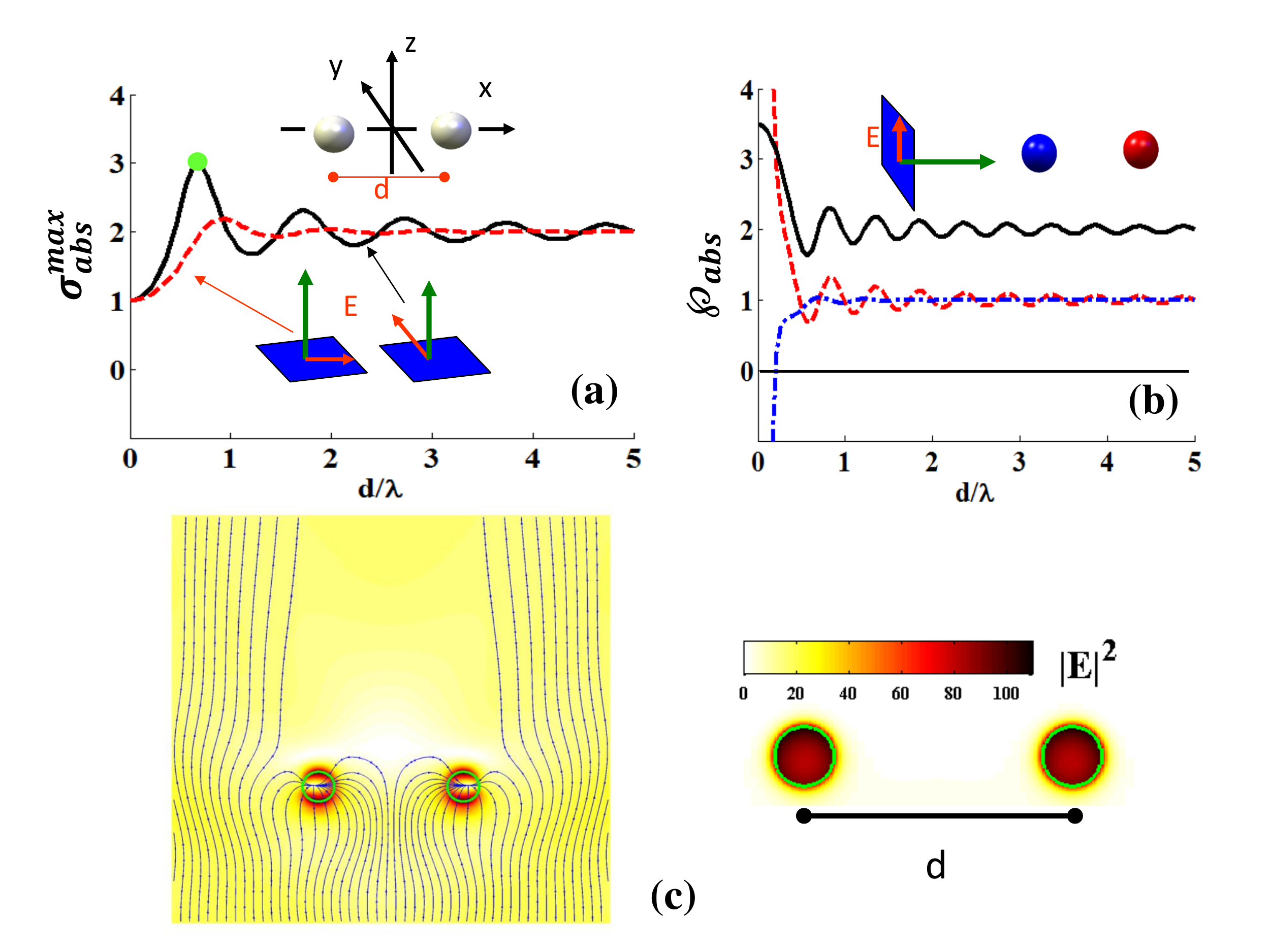}
\caption{Maximal absorption cross-section of a pair of electric dipoles enlightened (a) in a transversal direction (TE and TM polarization) and (b) in a longitudinal direction. In this case, the power dissipated by the first dipole is plotted in blue dot dashed line while the power dissipated by the second dipole is plotted in red dashed line. At subwavelength separation distance the first dipole is amplifying. All curves are normalized by the maximal cross-section of (resp. the power dissipated by) a single dipole.(c) Poynting vector streamlines around a dimer of SiC nanoparticles optimized to reach the maximal absorption marked by the green disk in Fig. 1(a) for a distance $d=7.3\mu m$ and a wavelength $\lambda=10.8\mu m$. The optimal radius is $0.805 \mu m$. The electric field inside the particles is two order larger than the unitary incident field.   } 
\end{figure} 

The above analyzis gives the ultimate limits that any dipolar system can reach given a spatial configuration. Surprisingly, these limits are independent of the material properties. However, hereafter we show that, in practical point of view, these values can be reached  by using appropriate nanoparticles. To this end, let us consider  the general relation $\mathbf p=\mathbf\alpha\mathbf{E}^{reg}_{loc}$ between the generalized polarizability of dipoles and the regularized version of local field defined as $\mathbf E^{reg}_{loc}(\mathbf{r}_i)=\mathbf E_{inc}(\mathbf{r}_i)+\omega^2 \mu_0\underset{j\neq i}{\sum}\mathds{G}_0(\mathbf{r}_i,\mathbf{r}_j)\mathbf p_{j}$  and let us assume, for clarity reasons, that all polarizability tensors are diagonal (i.e. $\mathbf\alpha=diag(\mathbf{\alpha_1},...,\mathbf{\alpha_N})$ with $\mathbf{\alpha_i}=diag(\alpha_{i,x},\alpha_{i,y},\alpha_{i,z})$). Then, by inverting this relation using the optimal dipolar moments we find for $i=1,...,N$ and $\beta=x,y,z$
\begin {equation}
\alpha_{i,\beta}=\frac{\mathbf p_{opt,i,\beta}}{\mathbf E_{inc,i,\beta}+\omega^2\mu_0\underset{j\neq i}{\sum}[\mathds{G}_{0}(\mathbf{r_i},\mathbf{r_j}).\mathbf p_{opt,j}]_{\beta}}.
\label{Eq:polarizability1}
\end{equation}
Here, we must emphasize that, in principle, those optimal polarizabilities do not necessary correspond to polarizabiliies of lossy media so that, generally speaking, the optimal absorption  can be achieved by combining dipoles with gain to lossy dipoles.

Let us now examine the usefull case of a  pair of dipoles (separation distance d) along the $\mathbf x$-axis: $Im\mathcal{G}=\left(\begin{array}{cc}
\frac{k}{6\pi}\mathds{1} & Im \mathds{G}_0(\mathbf{r_1},\mathbf{r_2}) \\
Im \mathds{G}_0(\mathbf{r_1},\mathbf{r_2})&\frac{k}{6\pi}\mathds{1}
\end{array}\right)$ and $Im \mathds{G}_0(\mathbf{r_1},\mathbf{r_2})=\left(\begin{array}{ccc}
a&0&0\\
0&b&0\\
0&0&b
\end{array}\right)$ with $a=\frac{sin(kd)}{2\pi d}\frac{1}{k^2d^2}-\frac{cos(kd)}{2\pi d}\frac{1}{kd}$ and $b=\frac{sin(kd)}{4\pi d}\frac{k^2d^2-1}{k^2d^2}+\frac{cos(kd)}{4\pi d}\frac{1}{kd}$ so that
$[Im\mathcal{G}]^{-1}=\left(\begin{array}{cc}
\mathcal{V} & \mathcal{W} \\
\mathcal{W}&\mathcal{V} 
\end{array}\right)$ 
with $\mathcal{V} =6\pi k\left(\begin{array}{ccc}
\frac{1}{k^2-(6\pi)^2 a^2} & 0 & 0 \\
0 & \frac{1}{k^2-(6\pi)^2 b^2} & 0 \\
0 & 0&\frac{1}{k^2-(6\pi)^2 b^2}
\end{array}\right)$
and $\mathcal{W} =-(6\pi)^2\left(\begin{array}{ccc}
\frac{a}{k^2-(6\pi)^2 a^2} & 0 & 0 \\
0 &\frac{b}{k^2-(6\pi)^2 b^2} & 0 \\
0 & 0&\frac{b}{k^2-(6\pi)^2 b^2}
\end{array}\right)$. For an incident field of magnitude $E_{inc}$ orthogonal to the axis linking the two dipoles, we have for TE waves  $\mathcal{P}^{max,\perp_{TE}}_{abs}=\frac{3}{2}\frac{\pi}{\omega\mu_0}\frac{E^2_{inc}}{k+6\pi b}$  while for TM waves $\mathcal{P}^{max,\perp_{TM}}_{abs}=\frac{3}{2}\frac{\pi}{\omega\mu_0}\frac{E^2_{inc}}{k+6\pi a}$). This result is plotted in Fig. 1-a. When the separation distance becomes sufficiently large then $b\sim a\sim 0$ and we see that the power absorbed by the pair is twice the power absorbed by an isolated dipole. On the contrary, close to the contact, $a\simeq\frac{1}{6\pi d}[kd-\frac{(kd)^3}{10}]$ and  $b\simeq\frac{1}{6\pi d}[kd-\frac{(kd)^3}{5}]$  so that $\mathcal{P}^{max,\perp_{TE,TM}}_{abs}=\mathcal{P}^{max,sing}_{abs}$. Between these two extreme regimes the maximum is an oscillating function with respect to the separation distance as illustrated in   Fig.1. in the case of a silicone carbide (SiC) nanoparticles dimer. In this case, the radius of SiC particles  has been optimized from the optimal  polarizability  (\ref{Eq:polarizability1}) using the  method described in \cite{Grigoriev} to reach the ultimate absorption shown by the green disk in Fig.1(a) at a separation distance $d=7.3\mu m$  for a wavelength $\lambda=10.8\mu m$ ($\epsilon_{SiC}=-2.53+0.18i$ \cite{Palik}). It clearly appears Fig. 1-c that the dissipation of the energy of incident field is due to a stong enhancement of field inside the particles.  
\begin{figure}[Hhbt]
\includegraphics[scale=0.34]{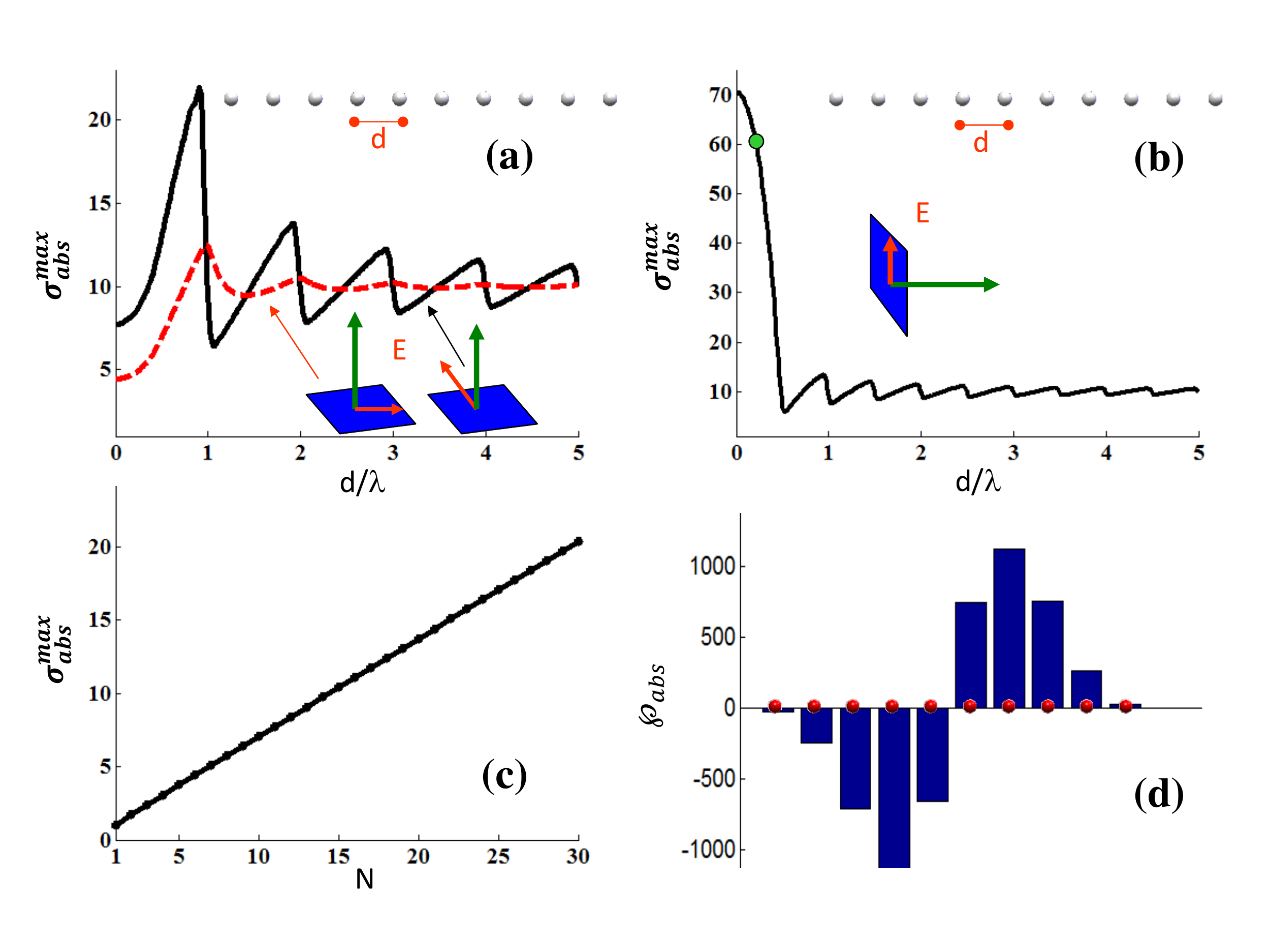}
\caption{Maximal absorption cross-section for chains of N=10 dipoles enlighted (a) in the transversal direction (TE and TM polarization) and (b) in the longitudinal direction. (c) Power dissipated per dipole inside optimal regular chain with different number N of dipoles separated by a distance $\frac{d}{\lambda}=0.25 $ for a longitudinal lighting (b). (d) Spatial distribution of losses and gains inside a chain of 10 dipoles ($\frac{d}{\lambda}=0.25 $) enlighted along its axis (green point in Fig. 2(b)). The cross-sections are normalized by the maximal cross-section of a single dipole. The power dissipated in each particle is normalized by the total power dissipated inside the chain.} 
\end{figure} 

When the dimer is lighted along its axis (see Fig. 1-b) both dipoles receive a phase shifted incident field. Let $\varphi=kd$ be that phase shift and  let us denote by $E_{inc}$ the magnitude of incident field on the first dipole. Then , it is straighforward to see that $\mathcal{P}^{max,_{//}}_{abs}=\frac{3}{2}E^2_{inc}\frac{\pi}{\omega\mu_0}\frac{k-6\pi b cos(\varphi)}{k^2-(6\pi)^2 b^2}$. In this case, at close separation distance  $\mathcal{P}^{max,_{//}}_{abs}=\frac{7}{2}\mathcal{P}^{max,sing}_{abs}$. This value is the maximal power a dimer can dissipate in this configuration. At first sight, this result seems to be paradoxal because one dipole is in the shadow of the second. But when we examinate the optimal losses per particle  in near-field regime (i.e. $\frac{2d}{\lambda}<1$) we see Fig. 1-b that one particle dissipates the incident energy while the second not. On the contrary, the latter is purely amplifiying. In this regime of strong coupling, the higher are the overall losses, the higher is the gain in the first particle. This exaltation of losses is driven by a cooperative coupling mechanism between a lossy and a particle with gain. Even when the first particle is lossless (at $\frac{d}{\lambda}\simeq 0.195$ in Fig. 1-b), the power which is dissipated in the second particle and therefore the overall  dissipated power is more than three time larger than the maximal power than a single isolated dipole can dissipate. As shown in Figs. 2  similar results can be observed with linear chains of dipoles.  The spatial distribution of losses inside such chains illuminated transversally and along the axis are plotted in Figs. 2-a and  2-b in the case of N=10 particles regularly spaced. In the longitudinal lighting case, we observe Fig. 2-c an almost symmetrical structure that emerges from the optimization process with half of the chain which is made of dissipative particles and the rest of the chain which is made of amplifying particles. As for the variation of the maximum absorption cross-section $\sigma_{abs}^{max}$ with respect to the number N of dipoles inside the chain,  the numerical simulations predict that it increases (Fig. 2-c) linearly as $\sigma_{abs}^{max}\sim\frac{2N+1}{3}$  showing so that apparently there is no upper bound for long chains provide had hoc optical properties are used.
  
In two dimensional networks (Fig. 3-a-and 3-c) we get qualitatively the same behaviors than in chains. In contrast, in three dimensional networks the optimal power is almost independent on the separation distance between the dipoles even in near-field regime where the electromagnetic couplings are strong. One can speculate that both the exaltation and inhibitation mechanims annihilate each other. However, the detailed investigation of three-dimensional dipolar networks goes far beyond the objective of the present work and it is in itsef a problem which would necessitate a specific study.  

It clearly appears in Fig. 3,  that the absorption cross-section of the whole system can, in near-field regime, be much larger than the apparent area of domain (shown by the green dashed lines) on which the dipoles are dispersed. The numerical simulations show that, at subwavelength separation distances, the cross-section can be more than one order of magnitude larger than the maximal cross-section of a simple dipole.This result outlines the importance of collective couplings in this regime and show its strong potential  for a variety of problem such as for instance, the design of subwavelength superabsorbers. Nevertheless, it is worth to emphasize that to get an absorption cross-section which is much larger than the apparent support of dipoles, active media  must generally be inserted inside the network. When we impose all dipoles to be dissipative we observe that the cross-section drastically decreases with the separation distance to become comparable with the maximal cross-section of a single dipole (shown by the horizontal blue dashed lined in Fig.3 ) near the contact.

\begin{figure}[Hhbt]
\includegraphics[scale=0.37, angle=0]{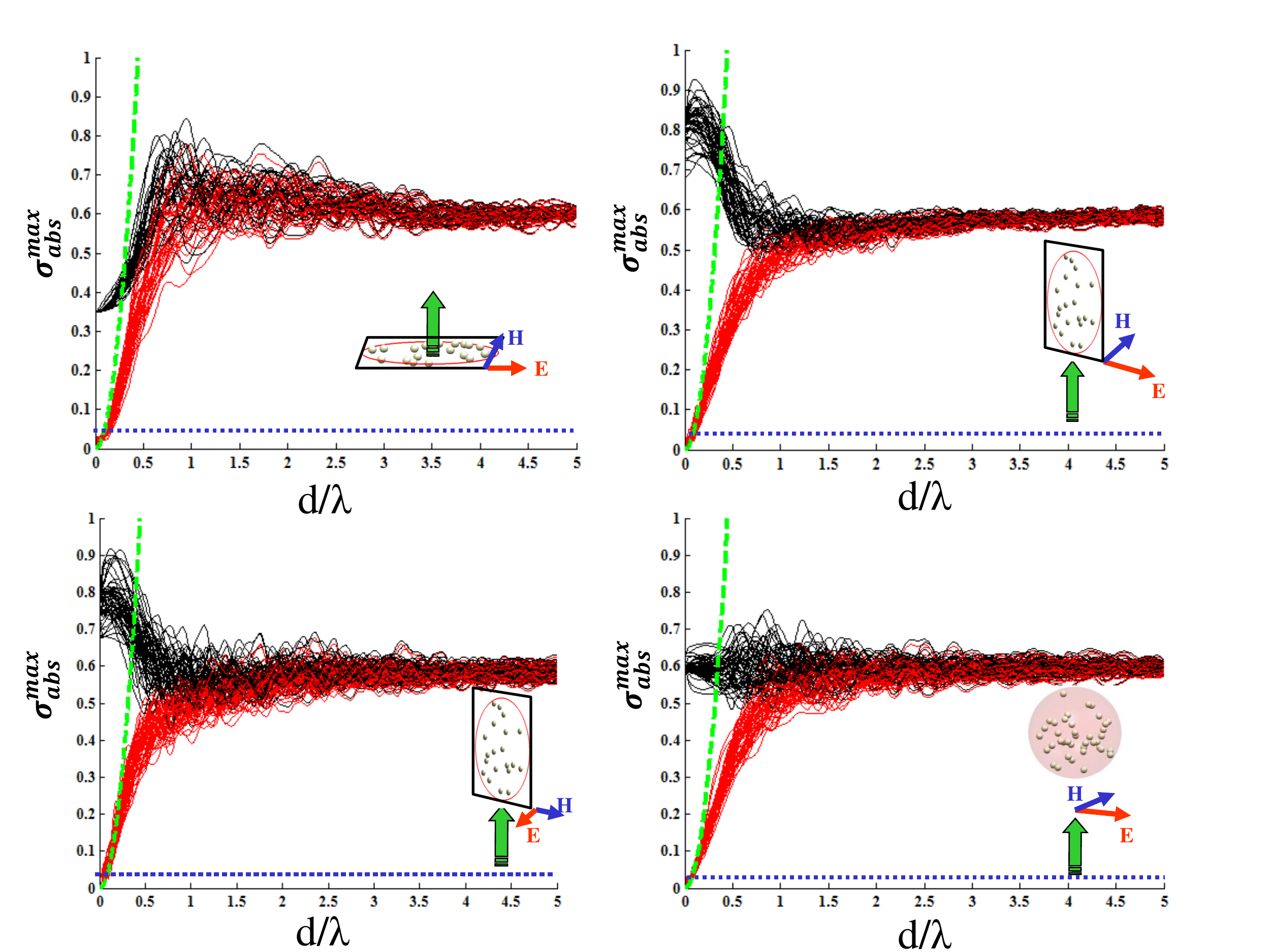}
\caption{ Maximal absorption cross-section of two-dimensional (resp. three dimensional) random networks of 20 dipoles and (resp. 40 dipoles) for a perpendicular and a longitudinal highligting with respect to the surfacic fraction in dipole.Here the results of 50 realizations (with a uniform probability density) are plotted for purely absorbing dipoles (red curves) and arbitrary dipoles (black curves).The blue dashed curve shows the maximal absorption cross-section for a single dipole. The green dashed curves denote the apparent area of domain in which the dipoles are distributed.} 
\end{figure}

So far, we have only considered couplings between particles described by simple dipoles. Hereafter we consider the most general situation where higher order modes  (ie. multipoles)  are taken into account \cite{Langlais}. The electromagntic field inside a medium of refractive index $n_h$ can be expressed in term of ingoing (-) and outgoing (+) vector spherical wave functions  (which form a complete basis)
\begin {equation} 
 \mathbf{\psi}_{pq}^{\pm}=\left(\begin{array}{c}
\mathbf{E}_{pq}^{\pm}\\
\mathbf{H}_{pq}^{\pm}
\end{array}\right) \label{Eq:spherical_wave}
\end{equation}
where we have adopted the usual convention for the multipolar index $(m,n)$ which are replaced by a single index $p=n(n+1)+m$ and where $q$ set the polarization state (i.e. $q=1$ for $TE$ waves and $q=2$ for TM waves). The outgoing wave functions $ \mathbf{\psi}_{pq}^{+}$ are solutions of Maxwell's equation (using the $e^{-i\omega t}$ convention)
\begin{equation}
\begin{cases}
 \nabla\times\mathbf{E}_{pq}^{+} =i\omega\mu\mathbf{H}_{pq}^{+}+\mathbf{H}_{pq}^{S}\\
 \nabla\times\mathbf{H}_{pq}^{+} =-i\omega\epsilon\mathbf{E}_{pq}^{+}+\mathbf{E}_{pq}^{S}
\end{cases}.\label{Eq:outgoing}
\end{equation}
with the source term $\mathbf{\psi}_{pq}^{S}=\left(\begin{array}{c}
\mathbf{E}_{pq}^{S}\\
\mathbf{H}_{pq}^{S}
\end{array}\right)$.
Now, let us introduce the following real spherical harmonic functions
\begin {equation}
\chi^{\pm}_{pq}=
\begin{cases}
\frac{1}{\sqrt 2}(\mathbf{\psi}_{pq}^{\pm}+\mathbf{\psi}_{pq}^{\pm})\; \text{if} \hspace{0.3cm} m<0 \\
\mathbf{\psi}_{pq}^{\pm}\hspace{1.8cm}  \text{if}\hspace{0.3cm} m=0\\
\frac{i}{\sqrt 2}(\mathbf{\psi}_{pq}^{\pm}-\mathbf{\psi}_{pq}^{\pm})\; \text{if} \hspace{0.3cm}m>0
\label{Eq.harmo_real}
\end{cases}
\end {equation}
and let us define the regular harmonics $\zeta_{pq}=\frac{\mathbf{\chi}_{pq}^{+}+\mathbf{\chi}_{pq}^{-}}{2}$. 
By definition, the incident field can be decomposed, in the complete basis of real spherical harmonic functions, as
\begin{equation}
\mathbf{\psi}_{inc}(\mathbf{r})=\underset{pq}{\sum} a_{inc,i}^{pq}\mathbf{\zeta}_{pq}(\mathbf{r}-\mathbf{r}_i). \label{Eq:field2}
\end{equation} 
As for the local field it  takes the form 
\begin{equation}
\mathbf{\psi}_{loc}(\mathbf{r})=\mathbf{\psi}_{inc}(\mathbf{r})+\underset{pq}{\sum} [a_{i}^{pq}\mathbf{\zeta}_{pq}^{(i)}(\mathbf{r}-\mathbf{r}_i)+ A_{i}^{pq}\mathbf{\chi}_{pq}^{+}(\mathbf{r}-\mathbf{r}_i)]. \label{Eq:field_local}
\end{equation} 
where the second term of rhs is the total scattered field which corresponds to  incoming field scattered  by all particles in  direction of the $i^{th}$ scatterer and the outgoing field diffracted out this particle. Since the incoming field is related to all outcoming fields by a linear relation of the type $\mathbf a_{i}^{pq}=\underset{j\neq i}{\sum}\underset{p'q'}{\sum}\mathds{T}_{ij}^{pq,p'q'}\mathbf A_{j}^{p'q'}$, where the $\mathds{T}_{ij}^{pq,p'q'}$ are the components of a translation operator (a propagator)  \cite{Stout} between the $i^{th}$ and the $j^{th}$ particle, the local field takes the form
\begin{equation}
\begin{array}{c}
\mathbf{\psi}_{loc}(\mathbf{r})=\hspace{7cm}\\
\underset{pq}{\sum} \mathbf{\chi}_{pq}^{+}(\mathbf{r}-\mathbf{r}_i)(\frac{1}{2}a_{inc,i}^{pq}+ \frac{1}{2}\underset{j\neq i}{\sum}\underset{p'q'}{\sum}\mathds{T}^{pq}_{ij} A^{pq,p'q'}_{j}+A_{i}^{pq})\\
+ \underset{pq}{\sum} \mathbf{\chi}_{pq}^{-}(\mathbf{r}-\mathbf{r}_i)(\frac{1}{2}a_{inc,i}^{pq}+ \frac{1}{2}\underset{j\neq i}{\sum}\underset{p'q'}{\sum}\mathds{T}^{pq,p'q'}_{ij} A^{p'q'}_{j}).
\end{array}
 \label{Eq:field_local2}
\end{equation}
Then, using the orthogonality relations for the functions $ \mathbf{\chi}_{pq}^{\pm}$, the power dissipated in each particle can be rewritten as the net flux
\begin{equation}
\begin{array}{c}\mathcal{P}_{abs,i}=\underset{pq}{\sum}\{-\mid \frac{1}{2} a_{inc,i}^{pq}+ \frac{1}{2}\underset{j\neq i}{\sum}\underset{p'q'}{\sum}\mathds{T}_{ij}^{pq,p'q'} \mathbf A_{j}^{p'q'}+\mathbf A_{i}^{pq}\mid^2\\
+\mid \frac{1}{2}\mathbf a_{inc,i}^{pq}+ \frac{1}{2}\underset{j\neq i}{\sum}\underset{p'q'}{\sum}\mathds{T}_{ij}^{pq,p'q'} \mathbf A_{j}^{p'q'}\mid^2\}
\end{array} \label{Eq:loss_general1}
\end{equation}
accross a surface surrounding the particle. By developping this expression we get
\begin{equation}
\begin{array}{c}
\mathcal{P}_{abs,i}=-\underset{pq}{\sum}\{\mid A_{i}^{pq}\mid^2+Re[A_{i}^{pq*} (a^{pq}_{inc,i})\\
+\underset{j\neq i}{\sum}\underset{p'q'}{\sum}T^{pq,p'q'}_{ij} A^{p'q'}_{j}]\}.
\end{array}\label{Eq:loss_general2}
\end{equation}
\begin{figure}[Hhbt]
\includegraphics[scale=0.43, angle=-90]{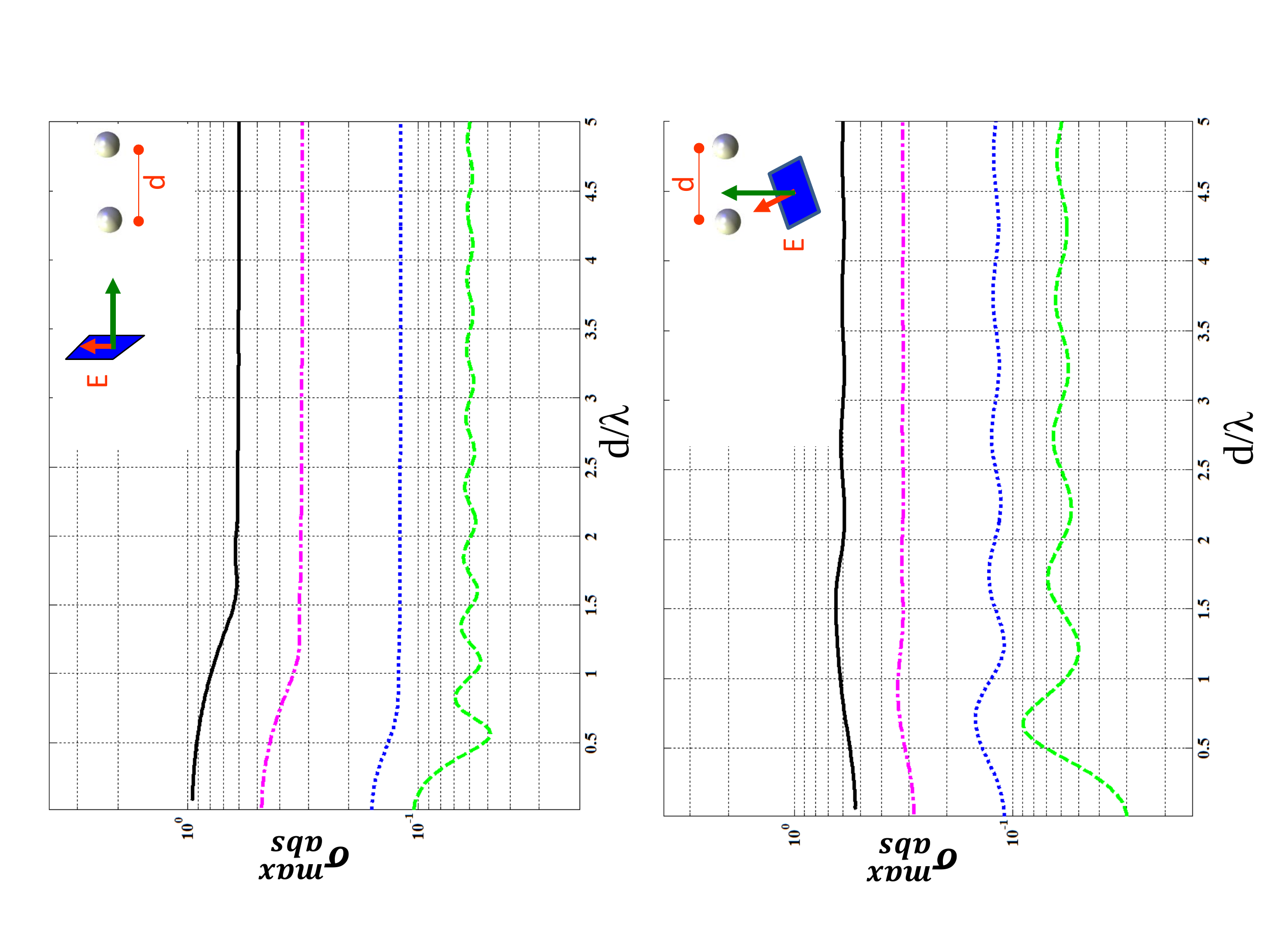}
\caption{Maximal absorption cross-section for a dimer of nanoparticles versus the separation distance.The green dashed line corresponds to the dipolar (electric) case while the blue, purple and black lines stand for the contribution of multipolar orders one (dipolar electric+magnetic), two and three, respectively.} 
\end{figure} 
Summing the losses over all dissipating objects we get, after a straightforward calculation, the total power dissipated by the system
\begin {equation}
\begin{array}{c}\mathcal{P}_{abs}(\omega)=\underset{ i}{\sum}\mathcal{P}_{abs,i}=-\mathbf A^\dagger.\mathbf V.\mathbf A+Re[\mathbf A^\dagger.\mathbf B]
\end{array}\label{Eq:power_abs_general4}
\end{equation} 
and the power scattered by it 
\begin {equation}
\begin{array}{c}\mathcal{P}_{scat}(\omega)=\mathbf A^\dagger.\mathbf V.\mathbf A.
\end{array}\label{Eq:power_scat_general4}
\end{equation} 
In these two expressions, $\mathbf A$ and $\mathbf B$ are, in a  system of N scatterers, the  block vectors defined with the subvectors $\mathbf A_i$ and 
$\mathbf{B}_i=-(\mathbf{a}^{pq}_{inc,i})$
while $\mathbf{V}=\mathds 1+\frac{\mathbf U+\mathbf U^\dagger}{2}$ is the block matrix defined with the sub-blocks elements $\mathbf U_{ij}=\mathbf T_{ij}$. Following the same reasoning as used for a set of dipoles we obtain the upper bounds for the dissipated and the scattered powers inside any arbitrary system of particles which are collectively coupled (see Supplemental Material \cite{SupplMat}).

These results are illustrated in Fig. 4 in the specific case of a dimer of nanoparticles. We note that, similarly to the results obtained for single objects \cite{Fan1}, the maximal cross-section is an increasing function with the number of multipolar orders. In fact the number of resonant modes of the whole system increases with the multipoles order creating so new channels for dissipating or scatter light.

In conclusion, we have analytically derived the ultimate limits for light absorption and sctattering by a system of N point resonant multipoles for a given geometry of their spatial distribution. We have demonstrated that these limits are independent of the optical properties of materials but they can  be reached only by combining lossy with active media. Beside the mechanisms of  coupling between light and single objects, the cooperative interactions  mechanisms in systems of coupled dissipating and amplifying resonant particles offer a supplementary degree of freedom to tailor light-matter interactions. Those mechanisms pave the way for promising scientific and practical applications in optics. 

\hspace{2cm}\textbf{ Supplemental material}

In this supplemental document we give supplementary informations concerning the derivation of the optimal power absorbed or scattered by a set of optical resonators. The derivation is general and can be used either in the dipolaror the multipolar case. 

\hspace{2cm}\textbf{I. Maximal absorption}

The maximal power which can be absorbed by such a system can be found by solving the following constrained optimal problem. Find  $\mathbf X_{opt}$,
\begin {equation}
\begin{cases}
\mathcal{P}_{abs}(\mathbf X_{opt})\rightarrow \underset{\mathbf X} {Max}\{\mathcal{P}_{abs}(\mathbf X) \}\\
{with}\hspace{0.5cm}  \mathcal{P}_{abs}(\mathbf X_{opt})\geq 0
\label{Eq.absorption}
\end{cases}
\end {equation}
where, according to relation (3) the absorbed power writes under the form $\mathcal{P}_{abs}=- \mathbf A^\dagger.\mathbf{V}.\mathbf A
+Re[\mathbf A^\dagger.\mathbf B]$ with $\mathbf A=\mathbf p$, $\mathbf{V}=\frac{\omega^3\mu_0}{ 2}[Im\mathcal{G}]$ and $\mathbf B=\frac{\omega }{2}\mathbf E_{inc}$ in the dipolar case. By decomposing this expression with respect to the real and imaginary parts of those vectors we have equivalently
\begin {equation}
\mathcal{P}_{abs}(\omega)=-\frac{1}{2} \mathbf X^\dagger.\mathcal{A}.\mathbf X+\mathcal{B}^\dagger.\mathbf X  
\label{Eq:power3}
\end{equation} 
with $\mathcal{A}=2\left(\begin{array}{cc}
\mathbf{V} & \mathbf{0} \\
\mathbf{0} & .\mathbf{V}
\end{array}\right)$,  
$\mathcal{B}=\left(\begin{array}{c}
Re[ \mathbf B] \\
Im[ \mathbf B]
\end{array}\right)$ and $\mathbf X=\left(\begin{array}{c}
Re[ \mathbf A]\\
Im[ \mathbf A]
\end{array}\right)$. 

\begin{figure}[Hhbt]
\includegraphics[scale=0.34]{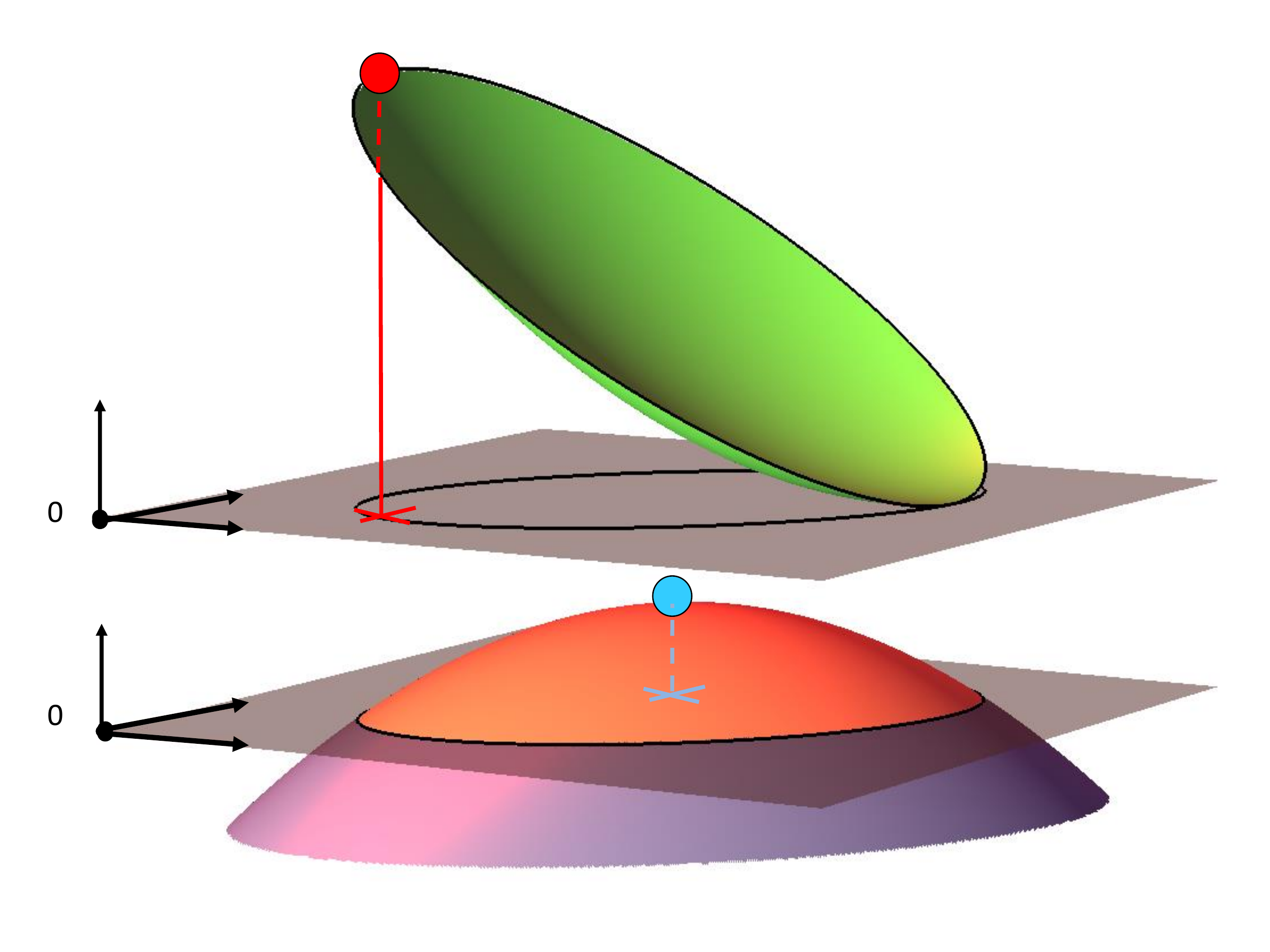}
\caption{Illustration of two optimization problems.On the bottom, the paraboloid corresponds to all physically admissible dissipated powers inside a system over the space of configuration (brown plane). The orange area on the upper part of this praboloid surface corresponds to lossy systems while the lower part correspond to amplifyiing systems. The blue sphere on the top of this surface denotes the maximal dissipated power while its orthogonal projection on the configuration plane stands  by a blue cross marks the optimal generalized polarizability. The upper  part of this mapping illustrates the maximization of the scattered power. The green paraboid  stands for all physically admissible scattered powers. The optimal generalized polarizability under the constraint of a positive dissipated power is shown by the red sphere.} 
\end{figure}

By construction $\mathcal{A}$ is a symmetric matrix so that expression (\ref{Eq:power3}) is a quadratic polynomial. Hence, the maximal dissipated power can be obtained by solving constrained optimal problem (1) using the classical Kuhn-Tucker (KT) conditions (see S. Boyd and L. Vandenberghe, Convex Optimization (Cambridge University Press, Cambridge, 2004). In the present case, this is equivalent to cancel the gradient $\frac{d\mathcal{P}_{abs}}{d\mathbf X}=-\mathcal{A}.\mathbf X+\mathcal{B}$ so that 
\begin{equation}
\begin{split}
 \mathbf X_{opt}=\mathcal{A}^{-1}.\mathcal{B}=\frac{1}{2}\left(\begin{array}{cc}
\mathbf V^{-1} & \mathbf{0} \\
\mathbf{0} & \mathbf V^{-1}
\end{array}\right).\left(\begin{array}{c}
Re[ \mathbf B] \\
 Im[ \mathbf B]
\end{array}\right)
\end{split}
\end{equation}
 It follows from the definition of vector $\mathbf X$ that the optimal generalized dipole vector field reads
\begin {equation}
 \mathbf A_{opt}=\frac{1}{2}\mathbf{V}^{-1}.\mathbf B.
\label{Eqopt_dipole}
\end {equation}
By inserting this expression into relation (3) of the Letter we get the maximal absorption
\begin {equation}
\mathcal{P}^{max}_{abs}(\omega)=\frac{1}{4} \mathbf B^\dagger.\mathbf{V}^{-1}.\mathbf B
\label{Eq:absorb_max2}
\end{equation} 
which is equivalent, accordng to the definition of different vectors and tensor fields to
\begin {equation}
\mathcal{P}^{max}_{abs}(\omega)=\frac{1}{8\omega\mu_0} \mathbf E_{inc}^\dagger.[Im\mathcal{G}]^{-1}.\mathbf E_{inc}.
\label{Eq:absorb_max2}
\end{equation} 

\hspace{2cm}\textbf{II.Maximal scattering}

 The maximal power which can be scattered by an arbitrary system can be found by solving the following constrained optimal problem. Find  $\mathbf X_{opt}$,
\begin {equation}
\begin{cases}
\mathcal{P}_{sct}(\mathbf X_{opt})\rightarrow \underset{\mathbf X} {Max}\{\mathcal{P}_{sct}(\mathbf X) \}\\
{with}\hspace{0.5cm}  \mathcal{P}_{abs}(\mathbf X_{opt})\geq 0
\label{Eq.scattering}
\end{cases}
\end {equation}
where $\mathcal{P}_{sct}=\mathbf A^\dagger.\mathbf V.\mathbf A$. Using the matrix $\mathcal{A}$ and the vector $\mathcal{B}$ introduced above the scatterd power writes equivalently $\mathcal{P}_{sct}=\frac{1}{2}\mathbf X^\dagger.\mathcal{A}.\mathbf X$ . Then, to solve the constrained optimal problem (6) we introduce the Lagrangian $\mathcal{L(\mathbf X,\lambda)}=\frac{1}{2}\mathbf X^\dagger.\mathcal{A}.\mathbf X+\lambda \mathcal{P}_{abs}(\mathbf X)$ where $\lambda>0$ denotes a positive  KT-multiplier \cite{Boyd}. Thus, according to the KT conditions, the optimal solution must cancel the gradient $\mathcal{L_{\mathbf X}}=\mathcal{A}.\mathbf X+\lambda(\mathcal{B}-\mathcal{A}.\mathbf X)$ of this Lagrangian.  This condition implies that  $\mathbf X=\frac{\lambda}{\lambda-1}\mathcal{A}^{-1}\mathcal{B}$. Beside, the constraint $\mathcal{P}_{abs}=0$ gives the optimal KT multiplier $\lambda=2$ so that  case $\mathbf X_{opt}=2\mathcal{A}^{-1}\mathcal{B}$. It follows from the definition of vector $\mathbf X$ that the optimal dipole vector field reads 
\begin {equation}
\mathbf A_{opt}=\mathbf {V}^{-1}.\mathbf B. 
\label{Eqopt_dipole2}
\end {equation}
Finally, by inserting this expression into the relation (6) of the Letter we get 
\begin {equation}
\mathcal{P}^{max}_{sct}= \mathbf B^\dagger.\mathbf{V}^{-1}.\mathbf B
\label{Eq:scatter_max2}
\end{equation}
which is equivalent to
\begin {equation}
\mathcal{P}^{max}_{sct}=\frac{1}{2\omega\mu_0} \mathbf E_{inc}^\dagger.[Im\mathcal{G}]^{-1}.\mathbf E_{inc}.
\label{Eq:scatter_max2}
\end{equation}

\textbf{III.A simple alternative for the derivation of optimum}

Here below we give another demonstration for the optimal powers absorbed or scattered by a N-body system following a purely algebraic approach. 
By construction $\mathbf V$ is a postive definite real matrix. Thus the two following scalars
\begin {equation}
Q_1= (\mathbf A^\dagger.-\frac{1}{2}\mathbf B^\dagger \mathbf V^{-1}).\mathbf V.(\mathbf A.-\frac{1}{2}\mathbf V^{-1}\mathbf B)
\label{Eq:scalar1}
\end {equation}
and
\begin {equation}
Q_2= (\mathbf A^\dagger.-\mathbf B^\dagger \mathbf V^{-1}).\mathbf V.(\mathbf A.-\mathbf V^{-1}\mathbf B)
\label{Eq:scalar2}
\end {equation}
are positive. From these expressions the absorbed and the scattered power given in Eqs.(3) and (6) of the Letter write
\begin{equation}
\mathcal{P}_{abs}=\frac{1}{4}\mathbf B^\dagger. \mathbf V^{-1}.\mathbf B-Q_1
\label{Eq.absorption3}
\end {equation}
and
\begin{equation}
\mathcal{P}_{scat}=\mathbf B^\dagger. \mathbf V^{-1}.\mathbf B-Q_2-2\mathcal{P}_{abs},
\label{Eq.scattering3}
\end {equation}
respectively. According to (\ref{Eq.absorption3}) $\mathcal{P}_{abs}$ is bounded by $\mathcal{P}^{max}_{abs}=\frac{1}{4}\mathbf B^\dagger. \mathbf V^{-1}.\mathbf B$
and from the definition of $Q_1$ we see that this value is reach when $ \mathbf A_{opt}=\frac{1}{2}\mathbf{V}^{-1}.\mathbf B$. Following a similar reasoning for the scattered power we have from (\ref{Eq.scattering3}) $\mathcal{P}^{max}_{scat}\leq\mathbf B^\dagger. \mathbf V^{-1}.\mathbf B$. Then, by choosing $\mathbf A=\mathbf V^{-1}\mathbf B$ we see, according to expression (\ref{Eq:scalar2}),  that $Q_2=0$ and $\mathcal{P}_{abs}=0$ and we reach the upper limit for the scattered power.

%
%



\begin{thebibliography}{38}
\bibitem{Nordlander1} J. A. Fan, C. H. Wu, K. Bao, J. M. Bao, R. Bardhan, N. J. Halas, V. N. Manoharan, P. Nordlander, G. Shvets, F. Capasso, Self-Assembled Plasmonic Nanoparticle Clusters, Science 328 (5982), 1135 (2010)
\bibitem{Nordlander2} J. B. Lassiter, H. Sobhani, J. A. Fan, J. Kundu, F. Capasso, P. Nordlander and N. J. Halas, Nano Lett., 10 (8), 3184 ( 2010)
\bibitem{Fano1} N. Liu, S. Kaiser and H. Giessen, Magnetoinductive and electroinductive coupling in plasmonic metamaterial molecules,  Adv. Mater., 20 (23) 4521 (2008).
\bibitem{Fano2} N. Liu, H. Liu, S. Zhu and H. Giessen, Stereometamaterials, Nature Photonics \textbf{3}, 157 (2009) 
\bibitem{EIT1} N. Liu, L. Langguth, T. Weiss, J. K\"{a}stel, M.Fleischhauer, T. Pfau and H. Giessen, Plasmonic analogue of electromagnetically induced transparency at the Drude damping limit, Nature Materials \textbf{8}, 758 (2009) 
\bibitem{EIT2} R. Taubert, M. Hentschel, J. K\"{a}stel and H. Giessen, Classical analog of electromagnetically induced absorption in plasmonics, Nano Lett., 12 (3), 1367 (2012)
\bibitem{Bohren} C. F. Bohren and D. R. Huffman, Absorption and scattering of light by small particles (John Wiley \& Sons, New York, 1983).
\bibitem{Fan1} Z. Ruan and S. Fan, "Superscattering of light from subwavelength nanostructures", Phys. Rev. Lett. \textbf{105}, 013901 (2010).
\bibitem{Fan2} Z. Ruan and S. Fan, "Design of subwavelength superscattering nanospheres",Appl. Phys. Lett. \textbf{98}, 043101 (2011).
\bibitem{Fan3} L. Verslegers, Z. Yu, Z. Ruan, P. B. Catrysse and S. Fan, "From electromagnetically induced transparency to superscattering with a single structure: a coupled-mode theory for doubly resonant structures", Phys. Rev. Lett. \textbf{108}, 083902 (2012).
\bibitem{Fleury} R. Fleury, J. Soric, and A. Al\`{u}, "Physical bounds on absorption and scattering for cloaked sensors", Phys. Rev. B \textbf{89}, 045122 (2014).
\bibitem{Alu} N. M. Estakhri and A. Al\`{u}, "Minimum-scattering superabsorbers", Phys. Rev. B \textbf{89}, 121416(R) (2014).
\bibitem{Grigoriev} V. Grigoriev, N. Bonod, J. Wenger and B. Stout, ACS Photonics, ACS Photonics DOI: 10.1021/ph500456w (2015).
\bibitem{Khokhlov} B. S. Luk'yanchuk, A. E. Miroshnichenko, M. I. Tribelsky, Y. S. Kivshar and A. R. Khokhlov, "Paradoxes in laser heating of plasmonic nanoparticles", New J. Phys. \textbf{14}, 093022 (2012).
\bibitem{Soljacic} W. Qiu, B. G. Delacy, S. G. Johnson, J. D. Joannopoulos and M. Soljacic, "Optimization of broadband optical response of multilayer nanospheres", Opt. Express, \textbf{20}, 18494 (2012).
\bibitem{Tretyakov} S. Tretyakov,"Maximizing absorption and scattering by dipole particles", Plasmonics, \textbf{9}, 935 (2014). 
\bibitem{Joannopoulos} O. D. Miller, C. W. Hsu, M. T. H. Reid, W. Qiu, B. G. DeLacy, J. D. Joannopoulos, M. Soljacic and S. G. Johnson, "Fundamental limits to extinction by mettalic nanoparticles",Phys. Rev. Lett. \textbf{112}, 123903 (2014).
\bibitem{Purcell} E. M. Purcell and C. R. Pennypacker, "Scattering and absorption of light by nonspherical dielectric grains," Astrophys. J. \textbf{186}, 705 (1973).
\bibitem{Draine} B. T. Draine and P. J. Flateau, "Discrete-dipole approximation for periodic targets: theory and tests," J. Opt. Soc. Am. A. \textbf{25}, 2693 (2008).
\bibitem{Jackson} J. D. Jackson, Classical Electrodynamics, third edition, John Wiley  (1999). 
\bibitem{Ben-Abdallah1} P. Ben-Abdallah, S.-A. Biehs, and K. Joulain, "Many-body radiative heat transfer theory," Phys. Rev. Lett. \textbf{107}, 114301 (2011). 
\bibitem{Ben-Abdallah2} R. Messina, M. Tschikin, S.-A. Biehs, and P. Ben-Abdallah, "Fluctuation-electrodynamic theory and dynamics of heat transfer in systems of multiple dipoles," Phys. Rev. B \textbf{88}, 104307 (2013).
\bibitem{Landau} L. Landau, E. Lifchitz, and L. Pitaevskii, Electromagnetics of Continuous Media (Pergamon, Oxford, 1984)
\bibitem{SupplMat} See EPAPS Document No. [number will be inserted by publisher] for a derivation of maximal power absorbed or scattered by a set of dipoles. For more information on EPAPS, see http://www.aip.org/pubservs/epaps.html.
\bibitem{Palik}\emph{Handbook of Optical Constants of Solids}, edited by E. Palik (Academic Press, New York, 1998).
\bibitem{Langlais} M. Langlais, J.-P. Hugonin, M. Besbes and Philippe Ben-Abdallah, Cooperative electromagnetic interactions between nanoparticles for solar energy harvesting, Optics Express, 22, S3, A577 (2014).
\bibitem{Stout} B. Stout, J.-C. Auuger, and J. Lafait, " A transfer matrix approach to local field calculations in multiple-scattering problems", J. of Modern Optics, \textbf{49}(13)  (2002).












\end{thebibliography}
\end{document}